# Electronic Structure of Copper Phthalocyanine: a Comparative Density Functional Theory Study

Noa Marom,[1] Oded Hod,[2] Gustavo E. Scuseria,[2] and Leeor Kronik[1*]

1. Department of Materials and Interfaces, Weizmann Institute of Science, Rehovoth 76100, Israel

2. Department of Chemistry, Rice University, Houston, Texas 77005

**Abstract**

We present a systematic density functional theory study of the electronic structure of copper phthalocyanine (CuPc), using several different (semi)-local and hybrid functionals, and compare the results to experimental photoemission data. We show that semi-local functionals fail qualitatively for CuPc, primarily because of under-binding of localized orbitals due to self-interaction errors. We discuss an appropriate choice of functional for studies of CuPc/metal interfaces and suggest the Heyd-Scuseria-Ernzerhof screened hybrid functional as a suitable compromise functional.

[*] Corresponding author. E-mail: leeor.kronik@weizmann.ac.il

## I. INTRODUCTION

In molecular solid form, copper phthalocyanine (CuPc) is a highly stable organic semiconductor with a broad range of applications in, e.g., light emitting diodes (usually as a hole injection layer), solar cells, gas sensors, thin film transistors, and even single molecule devices.[1] As is usually the case in electronics, performance of CuPc-based devices is often dominated by the properties of the CuPc interface with other organic and inorganic semiconductors, gate dielectrics, and metal electrodes. Hence, there is considerable interest in investigating the electronic structure of CuPc in general and the electronic structure of CuPc interfaces in particular. Many experimental studies (e.g., refs. 2-7) have been devoted to understanding interfaces of CuPc with various materials, from both the applied and the basic science point of view. Still, general trends of band alignment, band bending, formation of surface dipoles, charge transfer, potential barriers, and interface states, and their effect on charge transport and device performance,[8] are only partially understood.

Computational studies can provide a firm basis for the interpretation of experimental data and shed light on the underlying physics of such systems. Indeed, a number of first principles calculations for the electronic structure of CuPc, based on density functional theory (DFT), have been reported in the last decade.[9-16] However, as elaborated below, these studies yielded widely varying results.

One source of differences between the various calculations is the treatment of spin. Spin polarization in CuPc has been discussed thoroughly by Rosa and Baerends.[9] They have shown unequivocally that although copper is not usually associated with magnetic properties, the singly occupied 4s orbital of the Cu atom leads to significant spin-splitting and therefore CuPc must be treated in a spin-unrestricted manner. Indeed, most calculations have used an unrestricted spin configuration, but some have taken a spin-restricted approach, with significant differences ensuing, as discussed below.

A second source for major differences between the reported results, which is at the focus of this article, is the choice of the exchange-correlation functional. Some calculations[12,16] have used the local density approximation (LDA), in which the per-particle exchange-correlation energy at each point in space is approximated by that of a homogenous electron gas with the same local density; others[9-11,16] have used different flavors of the generalized gradient approximations (GGA), where deviations from homogeneity are accounted for by introducing density gradient corrections to the exchange-correlation energy. Others yet[13-15] have used the Becke 3-parameter Lee-Yang-Parr semi-empirical hybrid functional (B3LYP),[17] in which a fraction of exact (Hartree-Fock) exchange, as well as fractions of exchange and correlation gradient corrections, are mixed in an empirically determined manner.

While LDA and GGA perform satisfactorily for *some* organic molecules, B3LYP has become an almost default exchange-correlation functional for organic molecules in recent years.[18] However, B3LYP is not necessarily suitable for describing inorganic materials, because extended systems were not part of the data set against which B3LYP was parameterized. As recently discussed by Paier et al.,[19] because B3LYP uses components of Lee-Yang-Parr correlation, it is not exact in the limit of the uniform electron gas and does not contain a distinct treatment of opposite- and parallel-spin correlations, making it problematic for metals in general and for ferromagnetic metals in particular. Thus, B3LYP is not an obvious choice for understanding metal/CuPc interfaces.

A different hybrid functional where such difficulties are alleviated is PBEh[20] – a non-empirical one-parameter hybrid functional,[21] based on the GGA functional of Perdew, Burke, and Ernzerhof (PBE),[22] where the fraction of mixed in Fock-exchange is exactly 25% with no empirical parameterization. Because applying hybrid functionals to solids involves a high computational cost due to the long-range component of the Fock exchange,[21,23,24] it is also interesting to explore the recent screened hybrid of Heyd, Scuseria, and Ernzerhof (HSE).[23] In this PBEh-based functional, the Coulomb interaction is split into a long-range and a short-range component. The short-range component of the Fock exchange is mixed in just as in PBEh, but the long-range component is not, strongly reducing the computational cost.

In this paper, we systematically study the effect of a wide range of exchange-correlation functionals on the computed electronic structure of CuPc and compare the results to recent experimental data. We observe major qualitative differences between various functionals, elucidate their physical origin, and critically assess the pros and cons of using the various functionals studied for describing interfaces of CuPc with metals and semiconductors.

## II. COMPUTATIONAL DETAILS

All DFT calculations reported in this work were carried out using the Gaussian code.[25] 6-31G(d,p) basis sets were used for C, N, and H and the larger SDB-aug-cc-pVDZ basis set was used for Cu. The electronic structure of the CuPc molecule was studied using the following functionals: The Vosko, Wilk, and Nusair (VWN)[26] parameterization of LDA; two different GGA functionals: PBE[22] and BP86 (the latter a combination of Becke's 1988 exchange functional[27] and Perdew's 1986 correlation functional;[28] two different "conventional" hybrids: the semi-empirical B3LYP[17] and the non-empirical PBEh;[29,30] and the HSE screened hybrid.[23] Spin unrestricted calculations were employed throughout and the geometry was optimized independently for each functional.

## III. RESULTS AND DISCUSSION

The structure of the CuPc molecule is shown schematically in Fig. 1. CuPc is composed of a central Cu atom surrounded by four pyrrole rings, attached to benzene rings. The pyrrole rings are bridged by four additional N atoms. The molecule is planar with $D_{4h}$ symmetry. Bond lengths and angles obtained with the different functionals used in this work are listed in table 1. Generally, the choice of functional has no dramatic effect on the calculated geometry and the results are in agreement with previous calculations, e.g., those of ref. 31. Comparing to the hybrid functionals, the LDA functional yields slightly shorter bond lengths and the GGA functionals yield slightly longer bond lengths. This is typical of the behavior of these functionals for organic compounds.[18]

The Kohn-sham energy levels of CuPc, as calculated with the different functionals, are shown in Fig. 2. To facilitate comparison, the energy of the highest occupied molecular orbital (HOMO) was taken as zero throughout. Generally, the results can be grouped into those obtained from local and semi-local (namely, LDA and GGA) functionals and those obtained from hybrid functionals. One immediately apparent difference between the two groups is the gap between HOMO and the lowest unoccupied molecular orbital (LUMO), which is considerably smaller with the LDA/GGA functionals. The calculated HOMO-LUMO gap is 0.88 eV with LDA, 1.07 eV with PBE, 1.08 eV with BP86, 2.20 eV with B3LYP, 2.39 with PBEh, and 1.79 eV with HSE. At best, DFT-computed gaps are compared to the experimental optical gap (although there is no rigorous justification for such comparison[21]). Experiments conducted on CuPc thin films yielded optical gaps of 1.7 eV[32] and 1.5 eV.[33] These values are in closest agreement with the HSE-computed value, with significant underestimates and overestimates by the semi-local functionals and the conventional hybrids, respectively. The same trend was observed by Barone et al.[34] for semiconducting single-wall carbon nano-tubes. Note, however, that in addition to the assumption made in identifying the computed gap with the optical one, the optical gap of the thin film may differ from that measured in the gas phase.

Another obvious difference between the LDA/GGA functionals and the hybrids is that the LDA/GGA filled state spectra seem compressed with respect to the hybrid spectra, i.e., there is a general narrowing of the gaps between energy levels and more levels are "squeezed" into a given energy window (which can also be viewed as a higher density of states in these energy regions). To understand the possible origins of this, consider that Kohn-Sham eigenvalues are only approximations to quasi-particle excitation energies.[21] Hybertsen and Louie have shown that for many semiconductors and insulators, a direct comparison of (rigidly shifted) Kohn-Sham energy levels with quasi-particle excitation energies computed using many-body perturbation theory results in a fixed multiplicative "stretch factor" between the two.[35] Such stretching has also been observed in comparisons of LDA/GGA-computed spectra to experimental data for various organic systems.[36,37]

To determine whether such "stretching" is appropriate also for the present case, a comparison of density of states (DOS) curves (obtained from the computational data broadened by convolution with a Gaussian function) with the recent gas phase ultraviolet photoelectron spectroscopy (UPS) data of Evangelista et al.,[14] is given in Fig. 3 for both a higher and a lower experimental resolution. Clearly, the LDA/GGA spectra are

"compressed" also with respect to experiment, whereas the hybrid spectra are not. A comparison of appropriately[38] "stretched" LDA and GGA spectra is therefore also shown in Fig. 3. It is readily observed that this does not offer sufficient remedy. First, the multiplicative factors needed are very large (33% and 48% for LDA and PBE, respectively) and second, even after stretching, the lineshape of the LDA/GGA data is still in poor agreement with experiment.

In figure 3a, the calculated spectra, broadened by 0.4 eV wide Gaussian, are compared to the lower-resolution experimental data. The main four features of the experimental spectrum appear in all the calculated spectra, irrespectively of functional. However, a small satellite peak that appears on peak C in the hybrid spectra is found between peaks B and C in the LDA/GGA spectra, the former being in much better agreement with experiment. "Stretching" of the LDA/GGA spectra only accentuates the differences between them and the hybrid and experimental spectra: the shoulder of peak C is not reconstructed, the shape of peaks C and D are distorted, and the energy gap between peaks C and D is too large. A similar picture emerges in figure 3b, where the calculated spectra, broadened by 0.13 eV wide Gaussian, are compared to the higher-resolution experimental data. Peak A and what could be identified as the main four features of peak B (B1-B4 in the picture) appear in all spectra. However, in the LDA/GGA spectra, peak A has a satellite peak (instead of being sharply defined) and the energy gap between peaks A and B is considerably smaller than in the hybrid spectra and in experiment. Again, stretching obviously helps the gap between peak A and peak B, but after stretching, the satellite of peak A in the LDA and GGA data becomes a distinct peak with no obvious correspondence to experiment (this spurious peak should not be identified with the small experimental feature below the HOMO, which is known to be due to vibration replicas[15]). Interestingly, small differences between theory and the high-resolution experiment with respect to the width of peak B and in the precise position of the B sub-peaks persist even for the hybrid functionals. These could be due to effects not included in the calculation, such as final state effects and/or vibrational effect, or could reflect the residual inaccuracies of the functional. Note that in the experimental spectrum there is an additional feature, labeled F, which does not appear in any of the calculated spectra. It has been experimentally attributed to a Cu-derived state and previously suggested to disagree with DFT results due to final state effects.[14] Such effects are beyond the scope of the present text.

The inferior lineshape and extraneous peaks of the LDA/GGA spectra, even after "stretching", can be traced back to significant differences in orbital ordering between the eigenvalues obtained using the LDA, GGA, and hybrid functionals, as demonstrated in Fig. 4, which shows the orbitals associated with selected energies for selected functionals. For the occupied states, the most striking difference is in the description of the two $b_{1g}$ orbitals localized over the Cu atom and the surrounding pyrrole rings. The higher $b_{1g}$ orbital is spin-split and only its majority-spin orbital, which we denote as $b_{1g\uparrow}$, is occupied. In the LDA and GGA calculations the $b_{1g\uparrow}$ orbital is found at a considerably higher energy than in hybrid calculations. In GGA, at least the qualitative orbital ordering is retained, i.e., the HOMO orbital has $a_{1u}$ symmetry and the HOMO-1 is $b_{1g\uparrow}$. LDA fails qualitatively by predicting the $b_{1g\uparrow}$ orbital to be the HOMO and the $a_{1u}$ orbital to be the HOMO-1. Similarly, the lower $b_{1g}$ orbital, which is not spin-split, is also shifted to a much higher energy in LDA/GGA calculations, as compared to the hybrid calculations. With LDA/GGA, it is close in energy to the $e_g$, $a_{2u}$, and $b_{2u}$ orbitals, whereas in the hybrid calculations the lower $b_{1g}$ orbital is found at much lower energies.

The case of CuPc is reminiscent of that of another popular organic semiconductor molecule, 3,4,9,10-perylene tetracarboxylic acid dianhydride (PTCDA), where similar trends were found in the differences between the predictions of various functionals.[39] There, comparison to gas phase UPS data revealed that LDA and GGA strongly underestimate the binding energy of orbitals localized on the anhydride groups of the molecule, just as the binding energy of orbitals localized around the Cu atom is strongly underestimated in the present case. At the same time, for PTCDA the B3LYP result is consistent with both the experimental gas phase spectrum and the spectrum computed using many-body perturbation theory.[39] It is highly likely that, just as stipulated for PTCDA, this is due to the infamous self-interaction error (SIE),[21,40] i.e., the spurious Coulomb interaction of an electron with itself. SIE increases the Coulomb repulsion, causing a decrease in the binding energy and pushing upward the energy of the affected orbitals. The SIE is inherently larger for localized orbitals, such as the CuPc $b_{1g}$ orbitals, which are shifted to higher energies in LDA/GGA calculations, as compared to the hybrid calculations. Although hybrid functionals are not self-interaction free either,[21] the SIE for strongly localized orbitals is considerably alleviated by the partial inclusion of Fock exchange. This may explain the success of the hybrid calculations in predicting the energy levels of the occupied $b_{1g}$ orbitals of CuPc. Importantly, the HSE results are remarkably similar to those obtained from conventional hybrids. This is because the affected orbitals are strongly localized and therefore the long-range portion of Fock exchange, which is not included in HSE, has no significant effect on them. An additional piece of evidence which suggests that orbital ordering has to do primarily with exchange is that unreasonably large values of the Hubbard energy U were needed to correct the orbital ordering from a correlation point of view using the LDA+U approach.[16]

Additional important differences between LDA/GGA and hybrid functionals are found in the energies and ordering of the unoccupied states. The spin-splitting energy of the higher $b_{1g}$ orbital varies considerably according to the type of functional used in the calculation. It is smallest with LDA (0.88 eV), somewhat larger with GGA (1.32 eV for PBE, 1.36 eV for BP86), and much larger with the hybrid functionals (4.30 eV for B3LYP, 5.14 eV for PBEh, 4.32 eV for HSE). This difference in spin-splitting is much greater than the difference in gap values reported above and is therefore not merely a reflection of it. As a result of this difference in spin-splitting, the unoccupied minority spin orbital, $b_{1g\downarrow}$, is found at much lower energies in LDA/GGA calculations than in hybrid calculations. Consequently, in the LDA/GGA calculations the $b_{1g\downarrow}$ orbital is the LUMO, but in hybrid calculations it is the LUMO+1 and the LUMO is the doubly degenerate $e_g$ orbital. We interpret this as yet another manifestation of the SIE for $b_{1g}$ orbitals. Because the $b_{1g\uparrow}$ is spuriously shifted to higher energies, the $b_{1g\downarrow}$ must be spuriously shifted to lower energies in order to maintain the symmetry of the spin-splitting. As shown in figure 4, the magnitude of the spin-splitting of the $b_{1g}$ orbital determines the identity of the CuPc HOMO, HOMO-1, LUMO, and LUMO+1. In spin restricted calculations (not shown), this ordering is completely lost as the $b_{1g}$ orbital is found above the $a_{1u}$ orbital and below the unoccupied $e_g$ and becomes a singly occupied HOMO level.

Finally, we address the important question of the best "compromise functional" for investigating CuPc/metal interfaces The qualitative errors made by LDA/GGA for some of the most important orbitals of CuPc, as well as their overall gross quantitative failure, clearly indicates that we must strongly recommend against their use for CuPc. Therefore, they do not make for a reasonable compromise functional, which is disappointing given their excellent performance for many metals and semiconductors. Conversely, we have already stressed in the introduction that B3LYP is not a good compromise functional either, despite its excellent performance on the CuPc side, because it can perform quite poorly for some metals.[19] This leaves us with PBEh and HSE, both of which are, in principle, reasonable candidates. For organic molecules, HSE is known to yield results similar to those of PBEh.[41] In addition, PBEh and HSE do much better for metals than B3LYP.[19] Specifically, because HSE offers significant computational benefits for periodic systems, it is clearly preferable to PBEh. Its performance for solids is often comparable to that of GGA and at least remains qualitatively correct even when quantitatively not as accurate as GGA – see refs. 19, 42, and 43 for a detailed comparison and discussion. Thus, the reasonable performance of HSE for both the organic and inorganic sides of an interface, coupled with its low computation cost relative to conventional hybrids, leads us to recommend HSE as an attractive compromise functional for CuPc/inorganic interface, and likely other organic/inorganic interfaces as well.

## IV. CONCLUSION

We have calculated the electronic structure of CuPc using LDA, two flavors of GGA, conventional hybrids (both semi-empirical and non-empirical), and the screened-exchange HSE hybrid. All functionals describe the geometry of the molecule in a satisfactory manner; but differ greatly in the predicted electronic structure, including the assignment and energy position of some of the most chemically significant orbitals of CuPc. Comparing to hybrid functionals, the LDA and GGA functionals strongly underbind orbitals localized on the central region of the molecule and underestimate the spin-splitting of the $b_{1g}$ orbital, due to self interaction errors. As a result, the ordering of the orbitals is significantly altered and the identity of the HOMO and LUMO changes with the choice of functional. Although the spectra obtained with various functionals all superficially resemble experimental data, hybrid functional calculations clearly emerge as superior upon a more detailed analysis. However, periodic structures are more difficult to compute with conventional hybrid functionals and the results are typically less accurate than those of GGA. Thus, the HSE screened hybrid functional, which offers a qualitatively correct and quantitatively reasonable description of the electronic structure on both sides of an organic/inorganic interface, at a computational cost lower than those of conventional hybrids but still higher than that of semi-local functionals, emerges as a promising compromise functional.

## ACKNOWLEDGMENTS

Work at the Weizmann Institute was supported by the Gerhard Schmidt Minerva Center for Supra-Molecular Architecture. Work at Rice University was supported by the National Science Foundation CHE-0457030 and the Welch Foundation.

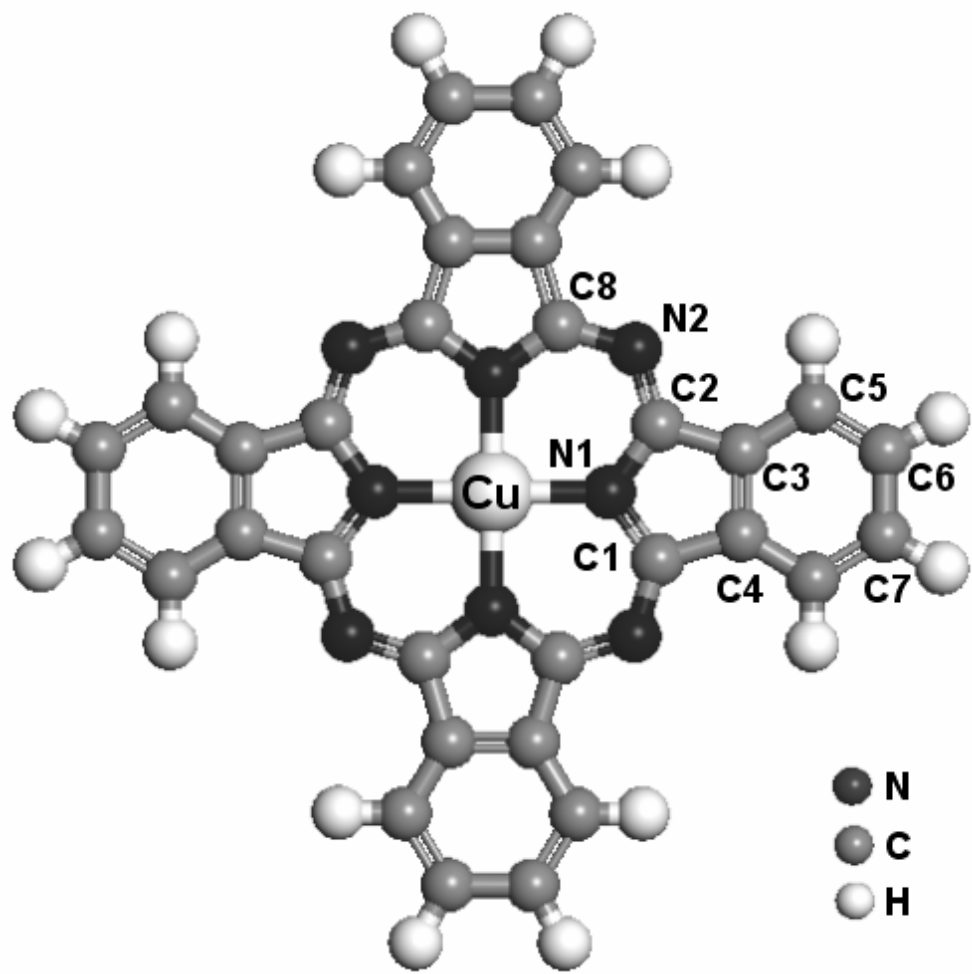

Fig. 1. Schematic view of the CuPc molecule

| CuPc Geometry | | | | | | | |
|---|---|---|---|---|---|---|---|
| | | VWN | PBE | BP86 | B3LYP | PBEh | HSE |
| Bond length [Å] | Cu –N1 | 1.935 | 1.970 | 1.969 | 1.968 | 1.956 | 1.956 |
| | N1 – C2 | 1.369 | 1.382 | 1.384 | 1.374 | 1.367 | 1.367 |
| | C2 – N2 | 1.317 | 1.332 | 1.333 | 1.326 | 1.321 | 1.321 |
| | C2 – C3 | 1.445 | 1.462 | 1.463 | 1.459 | 1.454 | 1.454 |
| | C3 – C4 | 1.400 | 1.415 | 1.416 | 1.407 | 1.402 | 1.402 |
| | C3 – C5 | 1.388 | 1.401 | 1.402 | 1.396 | 1.392 | 1.392 |
| | C5 – C6 | 1.390 | 1.401 | 1.402 | 1.394 | 1.390 | 1.390 |
| | C6 – C7 | 1.402 | 1.414 | 1.415 | 1.408 | 1.405 | 1.405 |
| Angle [°] | C2-N1-C1 | 107.986 | 108.602 | 108.478 | 108.795 | 108.678 | 108.694 |
| | N1-C2-N2 | 127.811 | 127.924 | 127.866 | 127.704 | 127.816 | 127.822 |
| | N1-C2-C3 | 109.620 | 109.203 | 109.267 | 109.147 | 109.299 | 109.279 |
| | C2-N2-C8 | 122.363 | 122.755 | 122.746 | 123.387 | 123.045 | 123.050 |
| | C2-C3-C4 | 106.387 | 106.493 | 106.494 | 106.456 | 106.362 | 106.374 |
| | C4-C3-C5 | 121.388 | 121.200 | 121.192 | 121.176 | 121.269 | 121.292 |
| | C3-C5-C6 | 117.326 | 117.587 | 117.596 | 117.633 | 117.484 | 117.494 |
| | C5-C6-C7 | 121.286 | 121.213 | 121.212 | 121.192 | 121.247 | 121.244 |

Table 1. Bond lengths and angles of CuPc, calculated with different exchange-correlation functionals

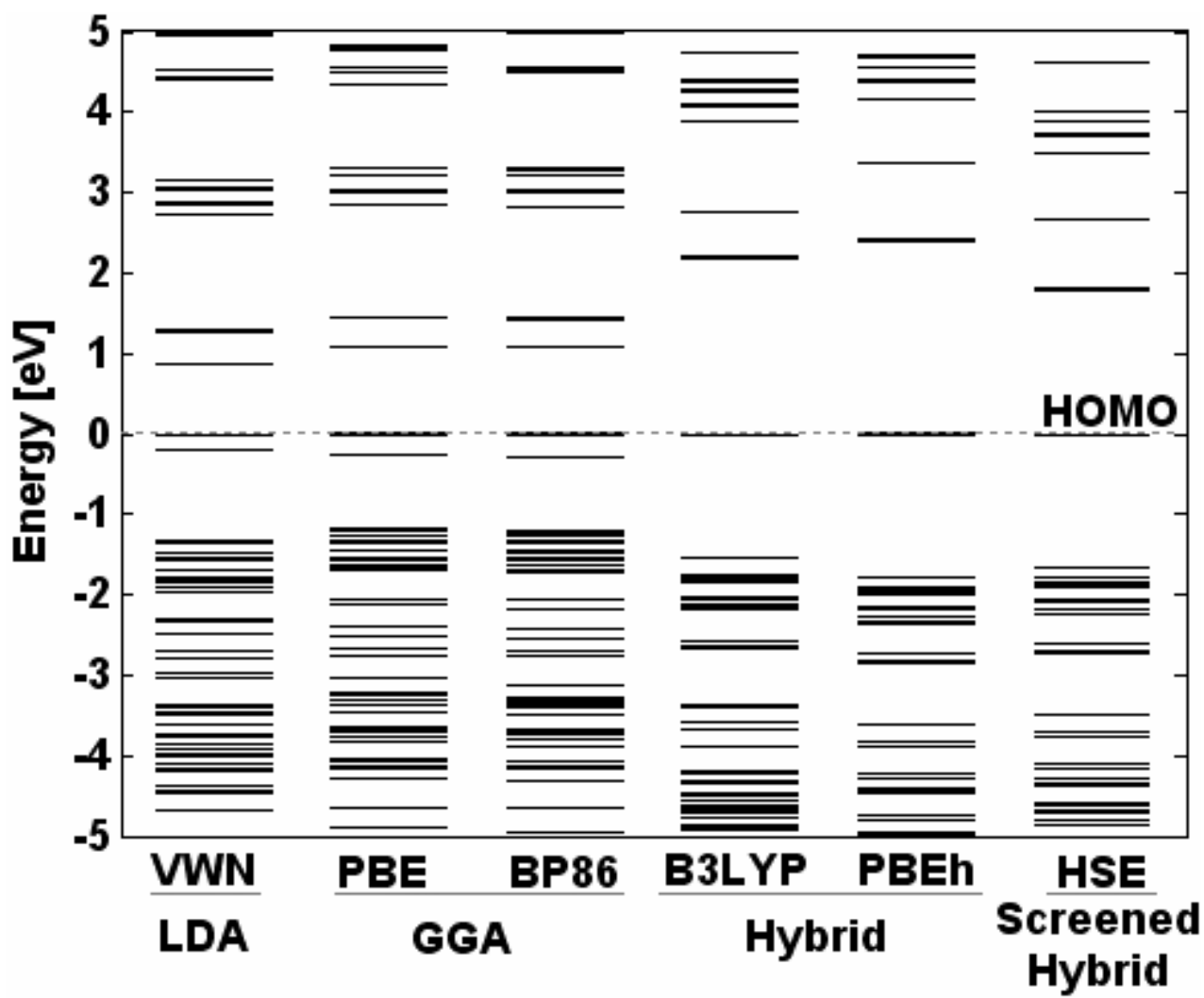

Fig. 2. Energy levels of CuPc calculated with different exchange-correlation functionals. All spectra have been shifted to align the highest occupied molecular orbital (HOMO).

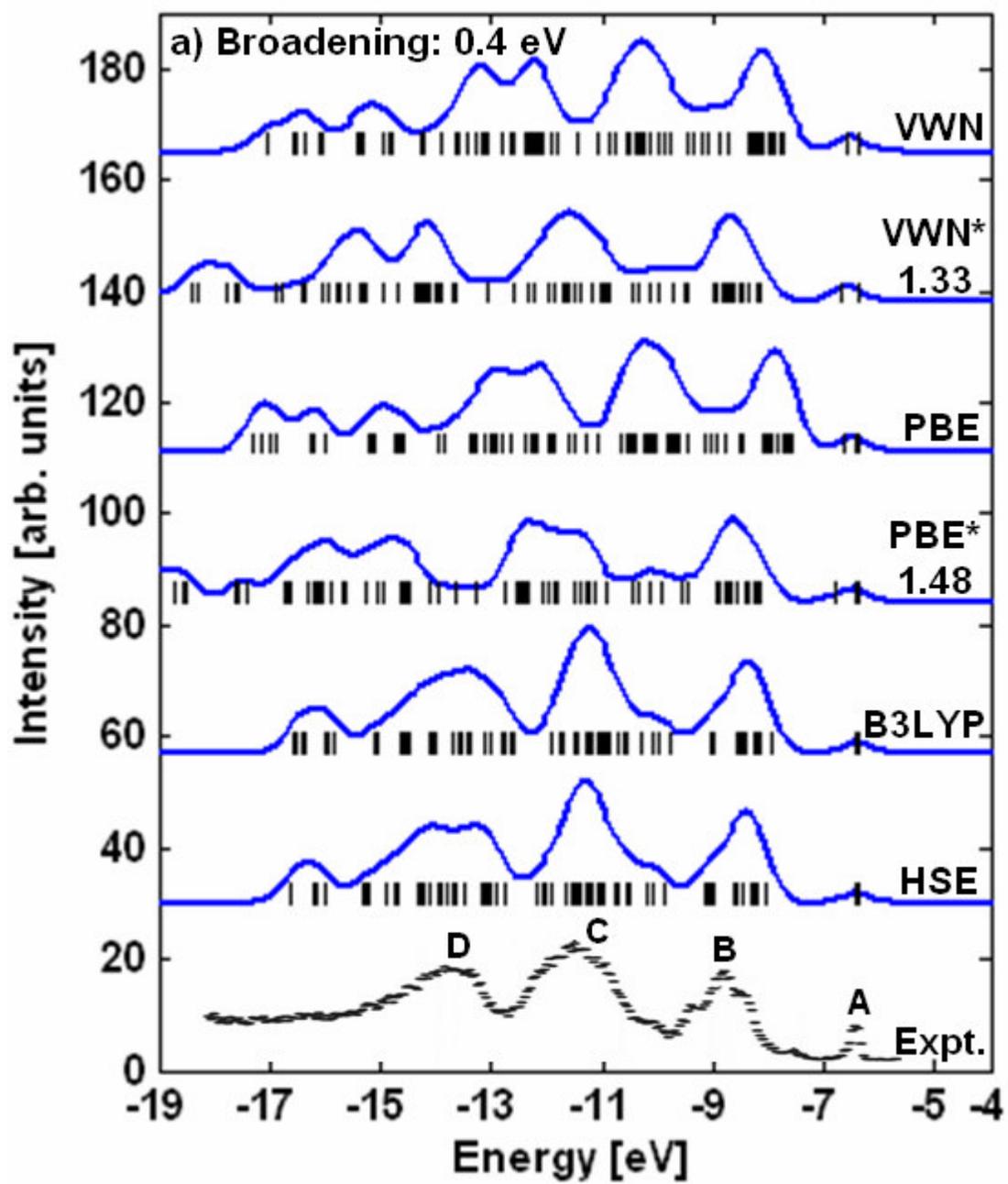

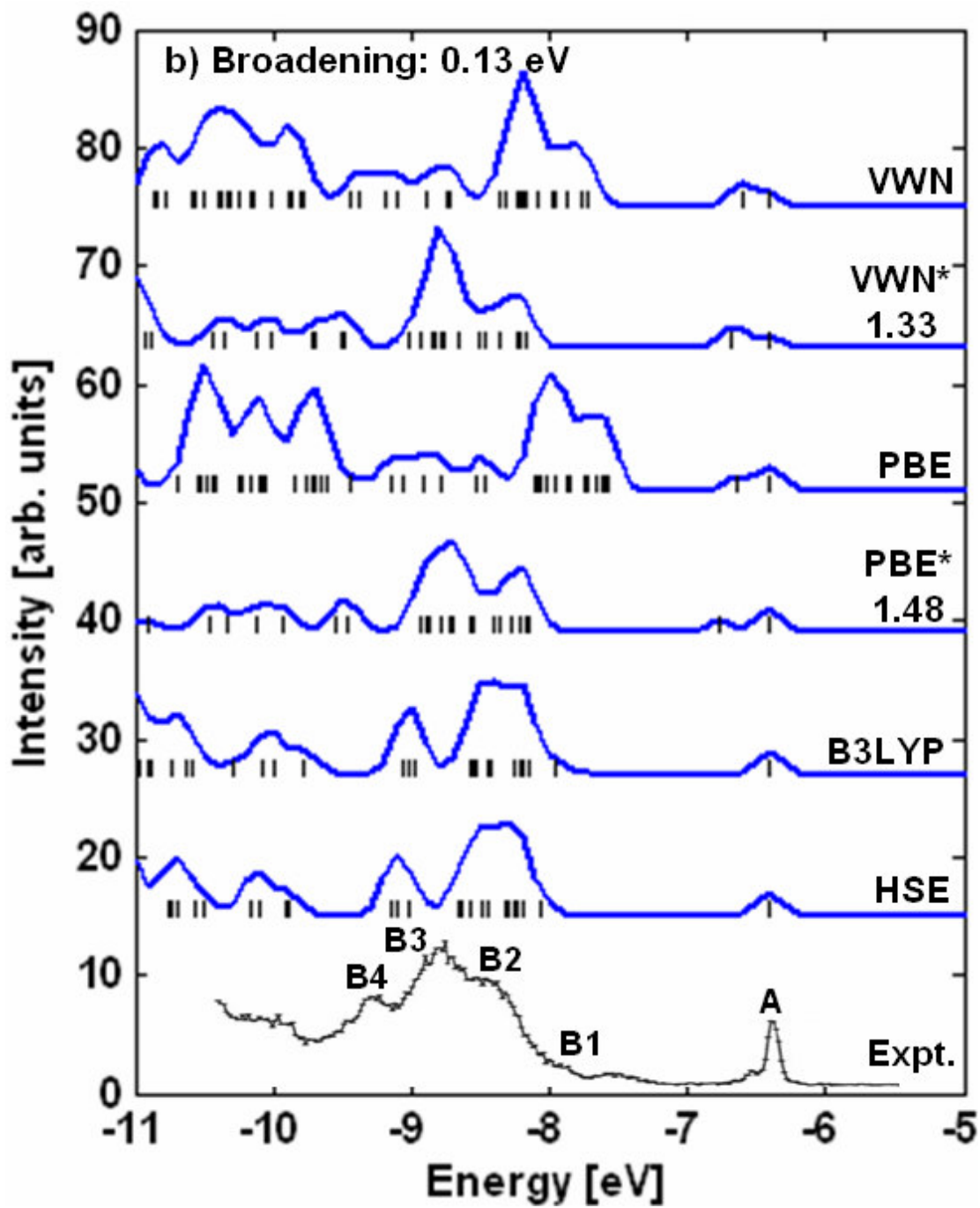

Fig. 3. CuPc spectra, calculated with different exchange-correlation functionals, compared to the gas phase UPS data of Evangelista et al.[14] a) calculated spectra, broadened by a 0.4 eV Gaussian, compared to the lower resolution experiment; b) calculated spectra, broadened by a 0.13 eV Gaussian, compared to the higher resolution experiment.

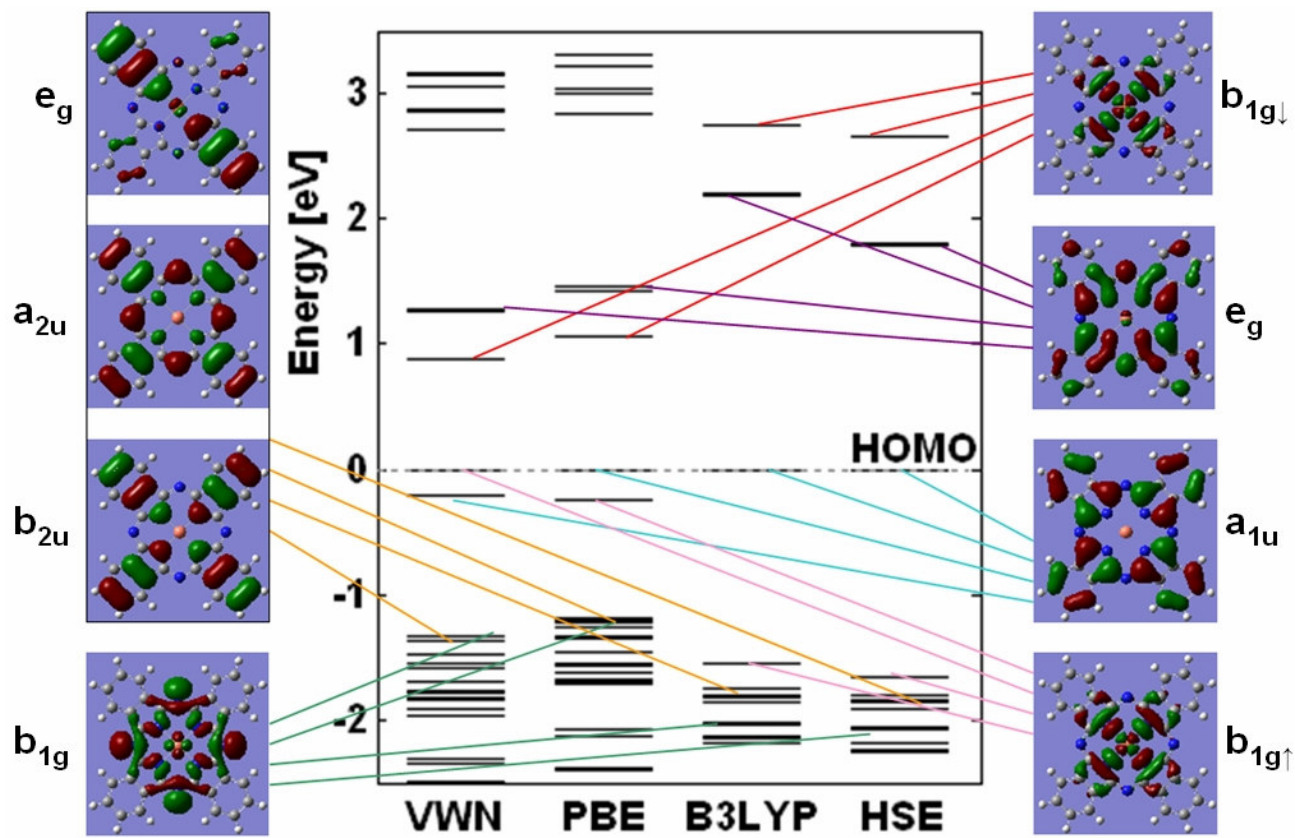

Fig. 4. Energy and ordering of selected CuPc molecular orbitals calculated with different exchange-correlation functionals. All spectra were shifted to align the highest occupied molecular orbital (HOMO). The $e_g$, $a_{2u}$, and $b_{2u}$ orbitals are very close in energy and are therefore denoted together. For clarity, only one example of each doubly degenerate $e_g$ orbital is shown.